\documentclass[final,5p,twocolumn,12pt]{elsarticle}

\usepackage{graphicx}
\usepackage{amssymb}
\usepackage{amsmath}
\usepackage{tikz}

\biboptions{sort&compress}

\usepackage{listings}
\input{listings-python.prf}
\lstnewenvironment{outputlisting}[1][]
  {\lstset{language=sh,#1}}{}

\lstnewenvironment{bashlisting}[1][]
  {\lstset{language=bash,#1}}{}

\lstnewenvironment{pythonlisting}[1][]
  {\lstset{language=Python,style=python-idle-code,#1}}{}

\usepackage{fancybox}
\newcommand{\code}[1]{\texttt{#1}}
\PassOptionsToPackage{normalem}{ulem}
\usepackage{ulem}
\usepackage{multirow}

\usepackage{colortbl}
\definecolor{lightgray}{gray}{0.8}

\newcounter{bla}

\usepackage{hyperref}

\journal{Computer Physics Communications}

\begin{document}

\newcommand{\BHM}{\begin{tt}BHM\end{tt}}
\newcommand{\FIXME}[1]{\textcolor{red}{#1}}

\begin{frontmatter}

\title{Implementation of the Bin Hierarchy Method for restoring a smooth function from a sampled histogram}

\author[a,b]{Olga Goulko\corref{author}}
\author[c]{Alexander Gaenko}
\author[c]{Emanuel Gull}
\author[a,d]{Nikolay Prokof'ev}
\author[a,d,e]{Boris Svistunov}

\cortext[author] {Corresponding author.\\\textit{E-mail address:} goulko@umass.edu}
\address[a]{Department of Physics, University of Massachusetts, Amherst, MA 01003, USA}
\address[b]{Present address: Raymond and Beverly Sackler School of Chemistry and School Physics and Astronomy, Tel Aviv University, Tel Aviv 6997801, Israel}
\address[c]{Department of Physics, University of Michigan, Ann Arbor, MI 48109, USA}
\address[d]{National Research Center ``Kurchatov Institute," 123182 Moscow, Russia}
\address[e]{Wilczek Quantum Center, School of Physics and Astronomy and T. D. Lee Institute, Shanghai Jiao Tong University, Shanghai 200240, China}

\begin{abstract}
We present {\BHM}, a tool for restoring a smooth function from a sampled histogram using the bin hierarchy method. The theoretical background of the method is presented in \cite{goulko2017bhm}. The code automatically generates a smooth polynomial spline with the minimal acceptable number of knots from the input data. It works universally for any sufficiently regular shaped distribution and any level of data quality, requiring almost no external parameter specification. It is particularly useful for large-scale numerical data analysis. This paper explains the details of the implementation and the use of the program.
\end{abstract}


\end{frontmatter}

{\bf PROGRAM SUMMARY}

\begin{small}
\noindent
{\em Manuscript Title: }   Implementation of the Bin Hierarchy Method for restoring a smooth function from a sampled histogram                      \\
{\em Authors: } Olga Goulko, Alexander Gaenko, Emanuel Gull, Nikolay Prokof'ev, Boris Svistunov    \\
{\em Program Title: }      BHM                                    \\
{\em Journal Reference:}                                      \\
{\em Catalogue identifier:}                                   \\
{\em Licensing provisions:}     GPLv3                               \\
{\em Programming language: C++}                                   \\
{\em Operating system:} Tested on Linux                                \\
{\em RAM:} 1--5~MB                                             \\
{\em Keywords:}  Data analysis, Function restoration, Spline fitting, Histogram, Smoothing\\
{\em Classification:} 4.9 \\
{\em External routines/libraries:} CMake, GSL \\
{\em Nature of problem:
} Restoring a smooth function from a sampled histogram in an efficient, reliable and automatized way is crucial for numerical and experimental data analysis. \\
{\em Solution method: }To make use of all information contained in the sampled data, the BHM algorithm generates a hierarchy of overlapping bins of different sizes from the initially supplied fine histogram. The bin hierarchy is fitted to a polynomial spline with the minimal acceptable number of knots, the positions of which are determined automatically. The output is a smooth function with error band. \\
{\em Running time: }Typically less than a second\\
\end{small}

\section{Introduction}
Numerical approaches to problems in condensed matter and quantum many-body physics often involve generating data points according to an unknown probability density $f(x)$, which needs to be restored from the sampled data. The amount of data generated in large-scale quantum Monte Carlo simulations is usually so large that it is impossible (or at least impractical) to store the complete list of sampled data points $x_i$, in order to use density estimation protocols \cite{KDEbookNarskyPorter, KDEbookScott, KDEbookSilverman} to recover $f(x)$. Instead, data points are typically collected into a histogram, the histogram bins representing integrals over the sampled distribution. This does not involve any significant loss of information, as long as the bins are sufficiently small to resolve the features of the distribution (which is always possible provided that $f(x)$ is sufficiently smooth). More sophisticated sampling methods exist, which retain more information about the individual points, but these are in general less efficient and require a case-dependent implementation. We provide a universal and efficient program to restore a smooth distribution, which uses the standard histogram as input.

{\BHM} is an implementation of the bin hierarchy method, introduced in \cite{goulko2017bhm}. 
It is
\begin{enumerate}
\item unbiased:
	\begin{itemize}
	\item utilizes all relevant information contained in the data;
	\item non-parametric fit automatically adjusts to data quality;
	\item provides maximally featureless solution (least acceptable number of spline knots);
	\end{itemize}
\item efficient:
	\begin{itemize}
	\item based on regular histogram, which is efficient to sample;
	\item fast analysis;
	\end{itemize}
\item automatic:
	\begin{itemize}
	\item very little user input;
	\item no adjustment for different types of sampled functions;
	\item no adjustment with simulation time as more data is collected.
	\end{itemize}
\end{enumerate}

The paper is organized as follows. The general problem setup is presented in Sec.~\ref{sec:problem}. In Sec.~\ref{sec:overview}, we give an overview of the algorithm. We explain how to use the program in Sec.~\ref{sec:io}, giving a detailed explanation of the input and output formats, as well as possible parameter specifications. Several examples are presented in Sec.~\ref{sec:examples}.

\section{Problem setup}
\label{sec:problem}
The central object in {\BHM} is a smooth function $f(x)$ defined on a bounded domain $\mathcal{D}$. Statistical sampling with a probability density $p(x)$ is performed to generate samples for $f(x)$ according to $f_j=f(x_j)/p(x_j)$ with $p$-distributed $x_j$. In the simplest case, when $f(x)$ itself is a normalized probability distribution, $p(x)=f(x)$ can be chosen, implying $f_j=1$. The samples are binned into a histogram with $2^K$ bins. We are interested in restoring a smooth function $\tilde{f}(x)$ from this histogram.

Each histogram bin $i$ with bin boundaries $x_{i,{\rm min}}$ and $x_{i,{\rm max}}$ represents the stochastic integral 
\begin{equation}
I_i=\int_{x_{i,{\rm min}}}^{x_{i,{\rm max}}} f(x) dx
\end{equation}
through the following relations:
\begin{eqnarray}
I_i&=&\bar{f}_i\frac{N_i}{N},\label{eq:I1}\\
M_2(I_i)&=&M_2(f_i) \, +\, \bar{f}_i^2 \, \frac{N_i(N-N_i)}{N},\label{eq:I2}\\
{\rm Var}(I_i)&=&\frac{M_2(I_i)}{N-1},\label{eq:I2b}\\
\delta I_i&=& \sqrt{\frac{{\rm Var}(I_i)}{N}},\label{eq:I3}
\end{eqnarray}
where the ``scaled variance'' $M_2(f_i)=(N_i-1){\rm Var}(f_i)$ is the sum of squares of differences from the mean, $N_i$ is the number of samples in bin $i$ and $N$ the total number of samples. Note that in the simplest case $p(x)=f(x)$ the above quantities are determined through $N_i$ and $N$ alone.

The goal is to find a function $\tilde{f}(x)$ whose \textit{integrals} over different parts of the domain $\mathcal{D}$ agree with the sampled integrals. Working with integrals rather than interpolated function values allows us to include combinations of histogram bins into the fitting. Rebinning data to larger bin sizes leads to a reduction of statistical noise, while retaining small bins results in a higher resolution due to smaller discretization errors.

The resulting fit $\tilde{f}(x)$ is a polynomial spline of order $m$, where $m$ is the highest power with non-zero coefficient. The spline function and its derivatives up to order $m-1$ are continuous, to account for the smoothness of the original $f(x)$.

\section{Overview of the algorithm}
\label{sec:overview}
\begin{figure*}[ht]
\begin{centering}
\includegraphics[width=\textwidth]{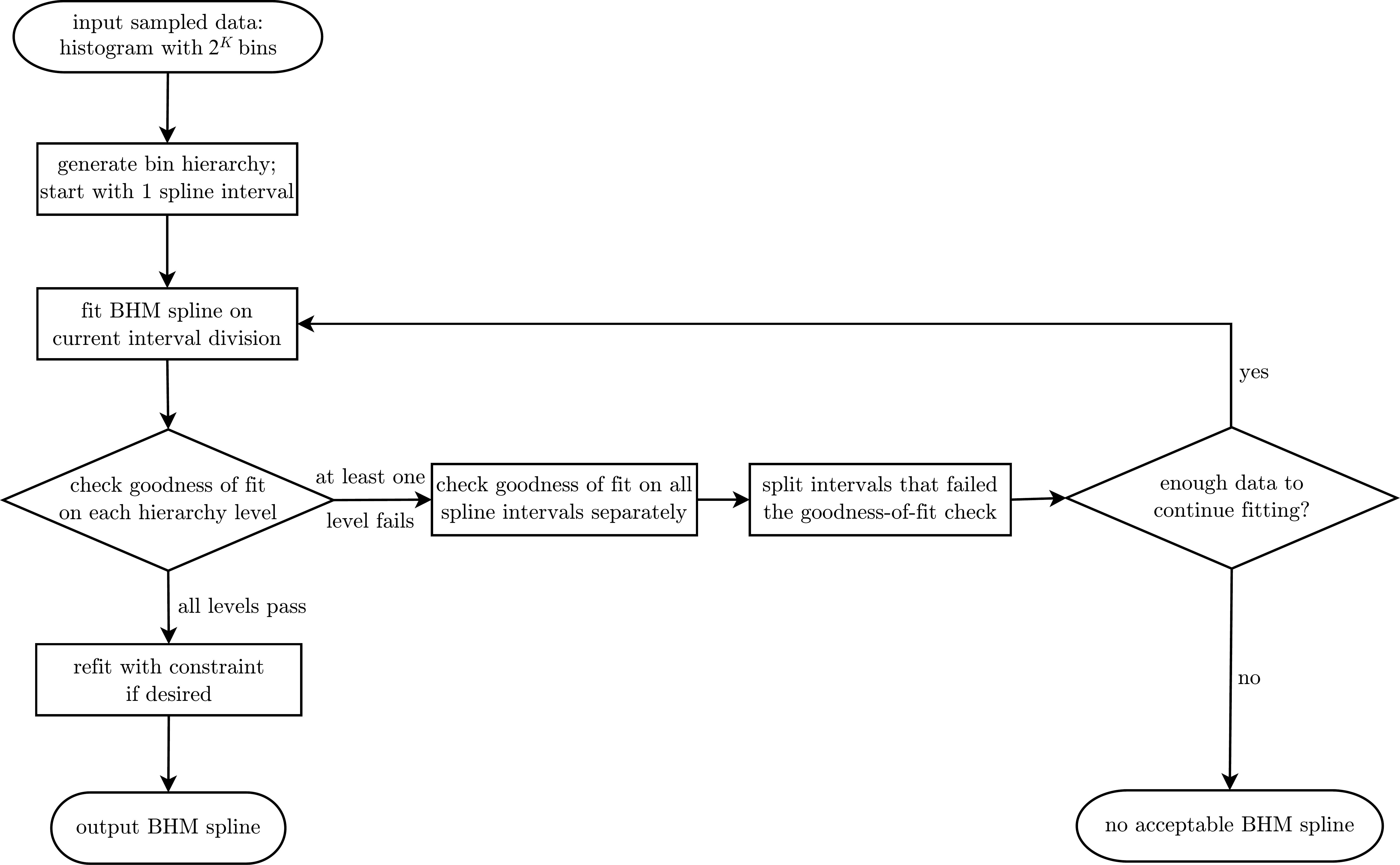} 
\caption{Flowchart of the algorithm\label{fig:flowchart}}
\end{centering}
\end{figure*}

In this section we give a brief overview of the algorithm. More details on the theoretical background of the method can be found in \cite{goulko2017bhm}. A flowchart of the algorithm is shown in Fig.~\ref{fig:flowchart}.

\begin{itemize}
\item The algorithm starts from a list of $2^K$ histogram bins supplied in an input file (for a detailed format description, see Sec.~\ref{sec:io}). Typical values of $K$ are 7 -- 15. It should be noted that the bins are not required to have the same size; however, in practice there is no need to have variable size bins. The bins must not overlap or leave gaps.
\item From this input the code generates a hierarchy of histogram bins of increasing size. Combining two neighboring bins of the $2^K$ initial bins leads to $2^{K-1}$ larger bins with, on average, twice as many entries. Successive repetitions of this rebinning result in a hierarchy of levels with $2^{K-2},\ldots,2,1$ bins on each level, respectively, where the final level consists of one bin over the entire domain containing an average over all sampled data. Bins that do not contain enough data for meaningful statistic, i.e.\ when the bin counter $N_i$ is smaller than a user defined minimal value, are excluded from the fitting process. Likewise, levels that do not contain enough usable bins (the minimal fraction can be defined by the user) are also excluded. This implies that in general fitting starts with a level $K'>K$ so that the original binning can be chosen to be very fine without introducing noise into the final fit. For bins that will be used for fitting, the bin integrals and their errors are computed via Eqs.~\eqref{eq:I1},\eqref{eq:I2}, \eqref{eq:I2b} and \eqref{eq:I3}.  
\item The code checks if the data is compatible with zero on the whole domain. There is an option not to proceed with the fit if this is the case. This feature is particularly useful for data suffering from a severe sign problem.
\item The next step is fitting a spline of order $m$ on the given spline interval division. The starting point is one spline interval, which means that one polynomial is fitted on the whole domain. The fit minimizes
\begin{equation}
\sum_{n=0}^K\frac{\chi^2_n}{2^n},
\label{eq:chisqmin}
\end{equation}
where $\chi^2_n$ is defined for bins on hierarchy level $n$ in the usual way.
\item Afterwards the goodness of fit is evaluated on each hierarchy level individually. The criterion is
\begin{equation}
\frac{\chi^2_n}{\tilde{n}}\leq 1+T\sqrt{\frac{2}{\tilde{n}}},
\end{equation}
where $T$ is the fit acceptance threshold (input parameter) and $\tilde{n}$ the number of bins on level $n$ that were used for fitting. The expression $\sqrt{2/\tilde{n}}$ corresponds to one standard deviation of the $\chi^2$-distribution.
\item If at least one level fails the global goodness-of-fit check, the goodness-of-fit is then evaluated on each spline interval separately (again level by level). Spline intervals on which the fit was acceptable remain unchanged, while the others are split into two parts, by introducing a spline knot in the middle (``number of bins"-wise).
\item If any of the resulting intervals is too small, meaning that there is not enough data to fit on that interval, the code exits without having produced an acceptable spline. Otherwise the BHM fit is repeated on the new interval division.
\item Once an acceptable spline has been found, there is an option to refit the data on the same interval division with an additional constraint that aims to minimize the jump in the highest derivative. 
\item The resulting BHM spline is output (spline coefficients and error coefficients). In addition, the spline values can be output evaluated on a grid.
\end{itemize}

\section{Input and output\label{sec:io}}
\subsection{Running the program}
Instructions for compiling the program and executing unit tests can be found in the README file.

The program executable requires 1 argument, the name of the parameter file, e.g.:
\begin{bashlisting}[numbers=none]
$ ./bhm in.param
\end{bashlisting}
In particular, the parameter file determines the name of the input file with the histogram data and the name of the output file for the BHM spline (see below).

As a special case, if the parameter file name is an empty string, the default parameters will be used which are suitable for most applications:
\begin{bashlisting}[numbers=none]
$ ./bhm "" <histogram.dat >spline.dat
\end{bashlisting}
In this case, the histogram data input is expected to be provided at the standard input, and the results will be printed to the standard output. In the example above, the standard input is redirected from file \verb|histogram.dat|, and the standard output is redirected to file \verb|spline.dat|.

Without an argument, the program prints a short help message and exits.

\subsection{Histogram input format}
The input histogram data is text-based, line-oriented, and has the following format:
\begin{outputlisting}
#[$A$] #[$N_{\mathrm{exc}}$]
#[$x_{1,\mathrm{min}}$]  #[$N_1$] #[$\bar{f}_1$] #[$M_2(f_1)$]
#[$x_{2,\mathrm{min}}$]  #[$N_2$] #[$\bar{f}_2$] #[$M_2(f_2)$]
....
#[$x_{i,\mathrm{min}}$]  #[$N_i$] #[$\bar{f}_i$] #[$M_2(f_i)$]
....
#[$x_{\mathrm{max}}$]
\end{outputlisting}
where the first line specifies an overall normalization factor $A$ and the number $N_{\mathrm{exc}}$ of samples outside of the histogram bounds. The normalization step is omitted if either $A=1$ or $A=0$. Otherwise, all values $\bar{f}_i$ and $M_2(f_i)$ are divided by $A$ and $A^2$, respectively, before constructing the BHM fit. The value $N_{\mathrm{exc}}$ is used to calculate the total number of samples $N=N_{\mathrm{exc}}+\sum_i N_i$, which is needed for Eqns.~(\ref{eq:I1}--\ref{eq:I3}). $N_{\mathrm{exc}}$ can be zero.

Starting from the second line, each line, except the last one, contains 2 or 4 blank-separated values, specifying the left bin boundary, the number of samples in the bin, and, optionally, mean value and scaled variance. For example, line~5 of the listing corresponds to a bin $i$ with the left boundary $x_{i,\mathrm{min}}$, number of samples $N_i$, mean value $\bar{f}_i$ and scaled variance $M_2(f_i)$ (see Eqns.~\ref{eq:I1}--\ref{eq:I3}). If the mean value and the scaled variance are both omitted, they are assumed to be $\bar{f}_i=1$ and $M_2(f_i)=0$, which corresponds to only ever adding 1 to bin counters, or in other words $p(x)=f(x)$. The last line of the file (line~7 of the listing) must contain a single entry $x_{\rm max}$, the right boundary of the last bin.

The numbers $x_{1,\mathrm{min}} \ldots < x_{i,\mathrm{min}} \ldots < x_{\mathrm{max}}$ must form a strictly monotonically increasing sequence, corresponding to non-overlapping, finite-size bins with no gaps. In the current implementation, the number of bins must be a power of 2 (in the later versions we may remove this limitation).

It is important to note that all sampled data and variances are assumed to be \textit{uncorrelated}. If correlations are present, they have to be removed prior to the BHM fit, for example through appropriate blocking analysis or by scaling the variances with the estimated correlation factor.

\subsection{Input parameter format}
The input parameter file is a text-based, line-oriented file that has a \code{key = value} format. An example input is shown in Fig.~\ref{fig:sample-input}. The keys are case-insensitive; the string values may be enclosed in quotes; the \verb|#| symbol starts a comment which is ignored until the end of the line. The meaning of each parameter is indicated in the figure in the corresponding comment. Below we provide more detailed explanations for some of the parameters.
\begin{figure*}[ht]
\begin{lstlisting}[language=Python,basicstyle={\small\ttfamily},frame=tb,title={Parameter File in.param},escapechar={!}]
DataPointsMin=100		#minimal number of data points per bin
SplineOrder=3			#spline order
MinLevel=2               	#minimal number of levels per interval
Threshold=2.0			#minimal goodness of fit threshold
ThresholdMax=4.0       		#maximal goodness of fit threshold
ThresholdSteps=4                #number of steps to take from Threshold to ThresholdMax
UsableBinFraction=0.25          #minimal proportion of good bins for a level to be used
JumpSuppression=false		#suppress highest order derivative (the error is unreliable)
Verbose=true		 	#verbose output
PrintFitInfo=true               #print the fit info
FailOnBadFit=true               #do not proceed if the fit is bad
FailOnZeroFit=true              #do not proceed if the data is consistent with zero
Data="histogram.dat"            #location of histogram input data
OutputName="spline.dat"         #location of the output file
GridOutput="spline_plot.dat"    #location of the spline-on-a-grid output
GridPoints=1024			#number of points for spline-on-a-grid output
\end{lstlisting}
\caption{\label{fig:sample-input}Sample parameter file}
\end{figure*}

\verb|DataPointsMin| in line~1 specifies the minimal number of data points that a bin must contain in order to be used for fitting. Bins that contain fewer sampled points are ignored (but still contribute in combination with other bins at higher hierarchy levels). \verb|DataPointsMin| must be at least 10, in order to ensure that meaningful statistics can be made from the data. The default value is 100. If a hierarchy level does not contain enough usable bins (the minimal number is given by the parameter \verb|UsableBinFraction| in line~7, times the total number of bins on that level) then this level and all subsequent levels are completely omitted from the fitting.

The maximal possible number of interval divisions is determined by the parameter \verb|MinLevel| in line~3. For example, if there are $2^K$ elementary bins, \verb|MinLevel=2| means that the smallest possible spline intervals coincide with the bins on hierarchy level $K-2$. \verb|MinLevel| must be at least 2 (corresponding to a total of at least 1+2+4=7 bins per interval); moreover, \verb|MinLevel| must be large enough to ensure that the fit is underdetermined for each interval.

The fit acceptance threshold $T$ (lines~4-6) can be either set to a fixed value, or to a range of values between \verb|Threshold| and \verb|ThresholdMax|. In the latter case, a BHM fit is first attempted with the smallest value \verb|Threshold|. If no acceptable fit is found, the threshold value is successively increased in \verb|ThresholdSteps| equidistant steps, until either an acceptable fit is produced or \verb|ThresholdMax| is reached. Setting \verb|ThresholdMax| to be smaller or equal to \verb|Threshold| and/or setting \verb|ThresholdSteps=0| corresponds to only using one fixed value of $T$. Note that threshold values that are too low can result in overfitting (too many spline pieces) or the failure to produce an acceptable fit. Values that are too high can result in underfitting (too few spline pieces and a poor fit with underestimated error bars). These issues are illustrated in Example~\ref{subsec:ex2}. The value $T=2.0$ is good generic choice. Specifying a range of threshold values reduces the statistical chance that there is no acceptable BHM fit with a given threshold, even though the data quality is adequate. The default range between $T=2.0$ and $T=4.0$ is suitable for most data sets.

Generally, the default parameter values in the example file are suitable for all types of sampled functions, and hence there is no need to change any of the parameters unless specifically desired.

\subsection{Output format}
The default verbose output is printed to standard error and contains auxiliary information such as values of the input parameters, a brief description of the input histogram, and the log of the fitting process. The fitting log is described in detail in Example~\ref{subsec:ex1}. If requested by the \code{PrintFitInfo} input parameter, information about the final fit is also printed to the standard output.

The output of the program is both human and machine-readable, and has the following text-based, line-oriented, blank-separated format:
\begin{outputlisting}
# Arbitrary comments
# ...
#[$m$] #[$s$]
#[$x_1$] #[$x_2$] ... #[$x_s$] 
# spline piece #[$1$]
#[$a_0$] #[$a_1$] #[$a_2$] ... #[$a_m$]
#[$\varepsilon_0$] #[$\varepsilon_1$] #[$\varepsilon_2$]... #[$\varepsilon_{2m}$] 
...
# spline piece #[$i$]
#[$a_0$] #[$a_1$] #[$a_2$] ... #[$a_m$]
#[$\varepsilon_0$] #[$\varepsilon_1$] #[$\varepsilon_2$] ... #[$\varepsilon_{2m}$]
# spline piece #[$(i+1)$]
...
\end{outputlisting}

Any lines at the beginning of the file that start with \verb|#| are considered comments and are ignored. The first significant line of the file (line~3 of the listing) specifies the spline polynomial order $m$ and the number of splines pieces $s$; the next line (line~4 of the listing) lists all $(s+1)$ spline piece boundaries $x_1,\ldots,x_{s+1}$. The following lines form $s$ sections describing each spline piece $\tilde{f}_i$, for $i=1\ldots s$. Each section (lines~5--7, 9--11 of the listing) consists of 3 lines:
\begin{enumerate}
\item Header (starts with \verb|#|) specifying the spline piece number ($i$),
\item $(m+1)$ numbers specifying the spline piece coefficients $a_0\dots a_{m}$ 
($\tilde{f}_i(x) = \sum_{k=0}^m a_k x^k$),
\item $(2m+1)$ numbers $\varepsilon_0\dots \varepsilon_{2m}$ specifying the error bar 
$E_i(x)=\sqrt{\sum_{k=0}^{2m} \varepsilon_{k} x^k}$.
\end{enumerate}

\subsection{Plotting the resulting spline\label{sec:plotting}}
The simplest way to plot the resulting spline is to use the provided Python3 script \verb|bhm_spline.py|, as follows:
\begin{bashlisting}[numbers=none]
$ python3 bhm_spline.py spline.dat  
\end{bashlisting}

On the other hand, it may be convenient to customize the plot and/or compare it with a known function, or plot it interactively (e.g., from a Jupyter notebook). For this purpose the script can be imported as a module that provides a BHM Spline class. The following listing demonstrates a possible way of using the module.

\begin{pythonlisting}[emph={bhm_spline,BHMSpline,spline},emphstyle=\color{blue}]
import numpy as np
import matplotlib.pyplot as plt
from bhm_spline import BHMSpline
    
spline=BHMSpline("spline.dat")
x=np.linspace(*spline.domain())
# reference function:
def fn(x): return (x**4-0.8*x*x)/0.171964
# plot the spline and the reference:
plt.plot(x,spline(x),  x,fn(x))
# plot the errorbar:
plt.plot(x,spline.errorbar(x))
# plot the spline with errorbars:
spline.plot()
# plot the spline and a reference:
spline.plot(fn)
# plot difference between spline and reference with error bar:
spline.plot_difference(fn)
\end{pythonlisting}

In line~3 the class \verb|BHMSpline| is imported; line~5 creates the object representing the spline. In line~6 an interval of x-values is created corresponding to the domain of the spline. Line~8 defines a reference function to compare with the spline. In line~10 the spline and the reference function are plotted using the \verb|Matplotlib| plotting library; in line~12 the error bar $E(x)$ is plotted. The class also provides a convenience plotting method: when called without arguments (as on line~14), the spline is plotted along with the error bars; when a function is passed as an argument (line~16), its graph is plotted also. It is also possible to plot the difference between the spline and the reference function with error bar (line~18).

\subsection{Grid output\label{sec:grid}}
If the \verb|GridOutput| parameter in the parameter file is set to a non-empty filename, the program also outputs to the specified file the values and the error bars of the spline computed on a one-dimensional grid of points. A plotting program, such as \verb|gnuplot|, can then be used to plot the generated function and the error bars and to compare them with a reference function; for example:
\begin{bashlisting}
$ gnuplot
gnuplot> quartic(x)=(x**4-0.8*x*x)/0.171964
gnuplot> plot "spline_plot.dat" with errors
gnuplot> replot quartic(x)
\end{bashlisting}
In this example, line~1 of the listing starts the \verb|gnuplot| program; line~2 defines a reference function (quartic polynomial); line~3 plots the grid output file generated by \verb|BHM|; and line~4 plots the reference function on the same graph.

\section{Examples}
\label{sec:examples}
In this section we present three detailed examples of the features of {\BHM} illustrated on different distributions $f(x)$. We provide a program to generate the input data for these examples (as well as for several additional test functions). Calling the program without arguments:
\begin{bashlisting}[numbers=none]
$ ./generator
\end{bashlisting}
prints a brief help message, which includes a list of the functions supported by the program.

Calling the program with a single file argument:
\begin{bashlisting}[numbers=none]
$ ./generator generator.param
\end{bashlisting}
generates the histogram data for a given analytical function according to the parameters listed in the \verb|generator.param| file. For all examples discussed below, the parameters are the same as shown in the example generator parameter file shown in Fig.~\ref{fig:generator-input} (including the random number generator seed), except when stated otherwise.

\begin{figure*}[ht]
\begin{lstlisting}[language=Python,basicstyle={\small\ttfamily},frame=tb,title={Parameter File generator.param},escapechar={!}]
SampleSize=10000		#total number of points sampled
Function=exponential		#name of test function
PowerBins=10			#generates uniform histogram with 2^PowerBins bins
RandomSeed=956475		#specify random seed (0 means random initialization)
Output="histogram.dat"		#histogram output file (missing means "standard output")
GridOutput="function.dat"	#output test function on a grid into this file (if provided)
GridPoints=1024			#number of grid points for test function output
\end{lstlisting}
\caption{\label{fig:generator-input}Sample parameter file to generate example input}
\end{figure*}

Calling the program as:
\begin{bashlisting}[numbers=none]
$ ./generator -python #[\textit{name}]
\end{bashlisting}
(where \textit{name} is the name of the function, possibly abbreviated) prints the Python code that corresponds to the function, which is convenient for plotting the analytical function against the approximating spline in an interactive Python environment (as has been discussed in subsection~\ref{sec:plotting}).

If the \verb|GridOutput| parameter in the parameter file is set to a non-empty filename, the program also outputs the values of the function computed on a one-dimensional grid to the specified file; a plotting program, such as \verb|gnuplot|, can then be used to plot the generated function; for example:
\begin{bashlisting}
$ gnuplot
gnuplot> plot "function.dat" with lines
gnuplot> replot "spline_plot.dat" with errors
\end{bashlisting}
In this example, line~1 of the listing starts the \verb|gnuplot| program; line~2 plots the generated function; and line~3 plots the content of the \verb|spline_plot.dat| generated by \verb|BHM| as discussed in subsection~\ref{sec:grid}.

\subsection{Example 1}
\label{subsec:ex1}
This example demonstrates BHM fits for different choices of spline order $m$.

The original function is a quartic polynomial (\texttt{Function=quartic\_polynomial}):
\begin{equation}
f(x)=\alpha(x^4-0.8x^2).
\end{equation}
Because $f(x)$ changes sign, sampling on the interval $[-1,1]$ is performed with the probability density $p(x)=|f(x)|$ and $\alpha=0.171964$ is chosen to ensure normalization of $p(x)$ on this interval.

\begin{figure*}[ht]
\begin{centering}
\includegraphics[width=\columnwidth]{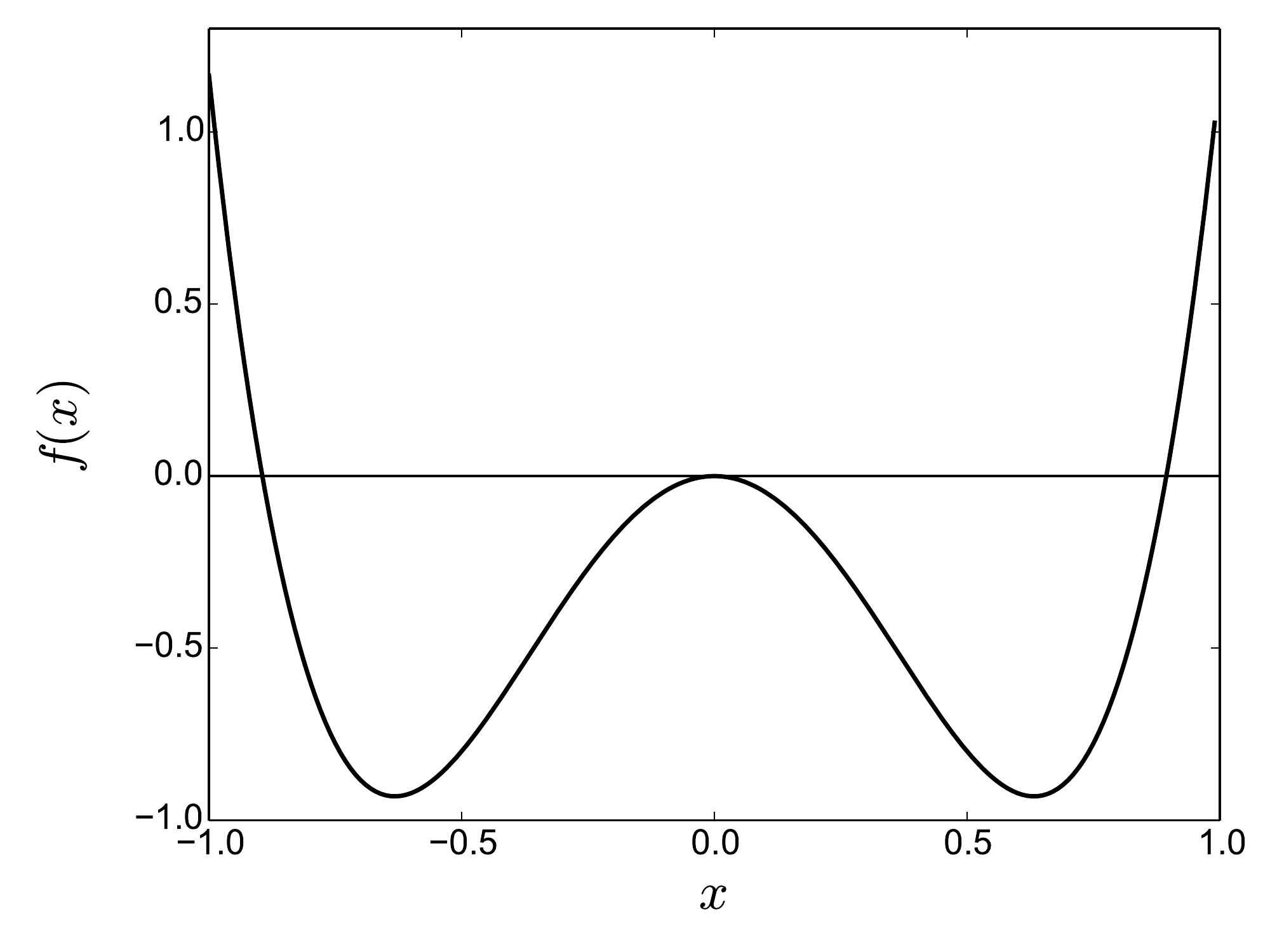} 
\includegraphics[width=\columnwidth]{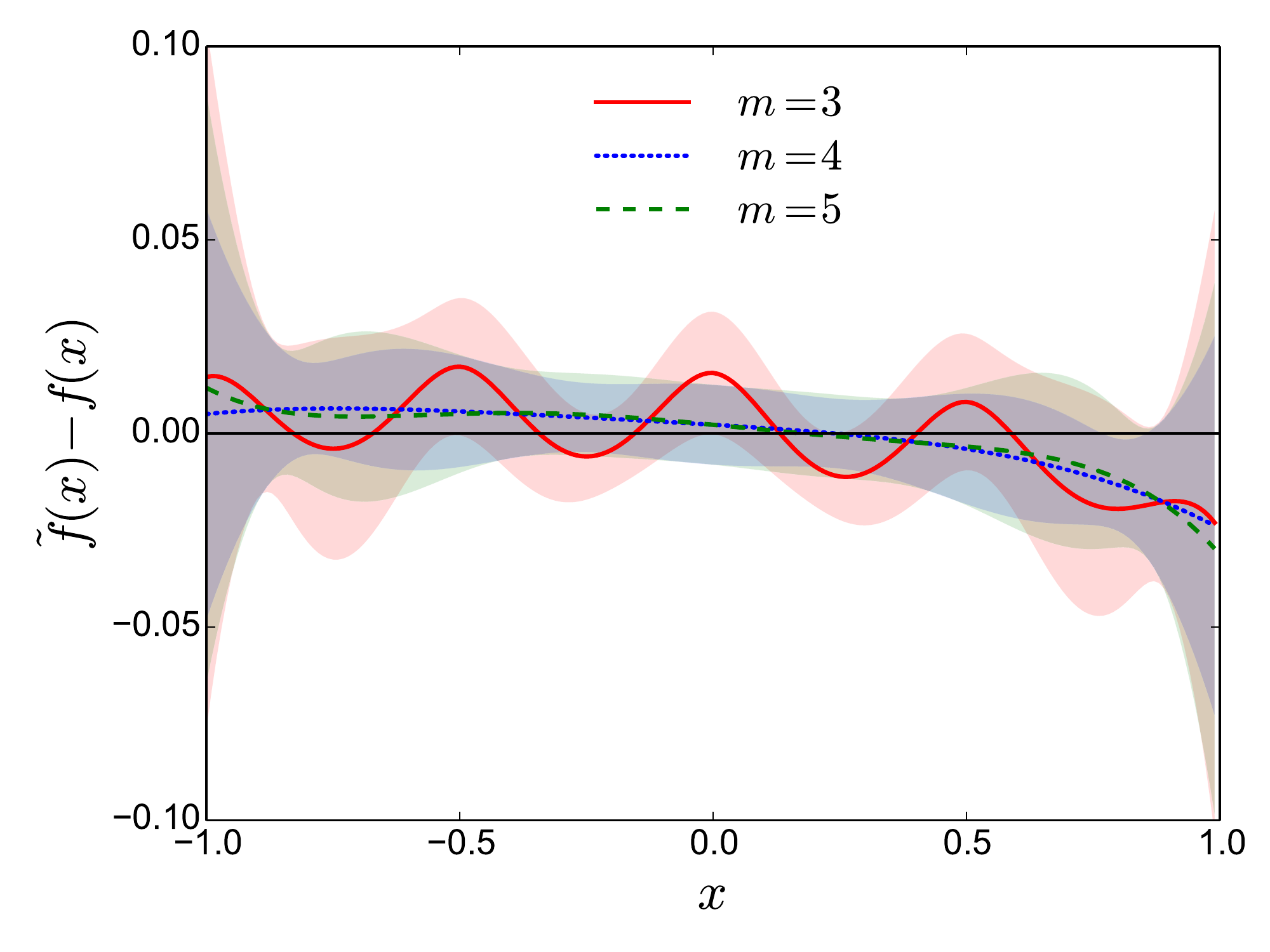} 
\caption{Quartic polynomial test function (left panel). Difference between BHM fit $\tilde{f}(x)$ with different spline orders $m$ and the test function $f(x)$ (right panel).\label{fig:ex1}}
\end{centering}
\end{figure*}

The histogram data is fitted with {\BHM} using the default parameters, with the exception of \texttt{SplineOrder} which is set to 3, 4, and 5 respectively. The fit results are shown in Fig.~\ref{fig:ex1}. From the output files \texttt{"spline.dat"} it can be seen that the cubic spline has four spline pieces; the quartic spline has one spline piece, as expected; the quintic spline also has one spline piece, its coefficients up to quartic order are similar to the ones obtained via quartic fit, and its highest spline coefficient is small.

We explain in detail the verbose output for the cubic fit $m=3$. At the beginning of the output, the fit parameters are listed, as well as general information about the input histogram. Then follows information about the goodness-of-fit at the different fitting stages:
\begin{outputlisting}
...
BHM fit:
Begin BHM fitting with threshold T = 2
Checking separate chi_n^2/n in spline fit
level   n       chi_n^2/n       max chi_n^2/n   
0       1          9.7585       3.8284          
1       2          2.7736       3.0000          
2       4          1.7636       2.4142          
3       8        832.1519       2.0000          
4       14       539.3412       1.7559          
5       24       210.2518       1.5774          
6       41       118.5452       1.4417          
7       54        67.7739       1.3849          
Checking interval 0 (order: 0, number: 0)
0       1          9.7585       3.8284          
This interval fit is not good
Checking separate chi_n^2/n in spline fit
level   n       chi_n^2/n       max chi_n^2/n   
0       1          0.0020       3.8284          
1       2          0.0006       3.0000          
2       4          1.6409       2.4142          
3       8          7.0923       2.0000          
4       14         7.8734       1.7559          
5       24         5.2920       1.5774          
6       41         3.3034       1.4417          
7       54         2.0987       1.3849          
Checking interval 0 (order: 1, number: 0)
1       1          0.0005       3.8284          
2       2          1.7172       3.0000          
3       4          6.3727       2.4142          
This interval fit is not good
Checking interval 1 (order: 1, number: 1)
1       1          0.0006       3.8284          
2       2          1.5645       3.0000          
3       4          7.8119       2.4142          
This interval fit is not good
Checking separate chi_n^2/n in spline fit
level   n       chi_n^2/n       max chi_n^2/n   
0       1          0.0001       3.8284          
1       2          0.0002       3.0000          
2       4          0.0055       2.4142          
3       8          0.0519       2.0000          
4       14         0.3837       1.7559          
5       24         0.8437       1.5774          
6       41         0.8378       1.4417          
7       54         0.8693       1.3849          
Good spline found with threshold T = 2
...
\end{outputlisting}
First a fit is attempted with one spline piece on the whole domain (lines~4-13). This fit is not acceptable because $\chi^2_n/\tilde{n}$ (third column in the output) exceeds the maximally allowed value $1+T\sqrt{2/\tilde{n}}$ (fourth column in the output) for most of the levels. The second column lists $\tilde{n}$, the number of available bins at each level. This number is in general smaller than $2^n$, because some bins do not contain enough data to be used for fitting. Also, hierarchy levels below $n=7$ were omitted because the fraction of usable bins on these levels was below the set \verb|UsableBinFraction| value.

Since the first fit was unsuccessful, $\chi^2$ is evaluated on each spline interval separately (lines~14-16). In this case, this yields no new information, since only one interval is present. As soon as a level is found where the fit is unacceptable (level~0 in this case), this check stops without proceeding to lower levels, since this is enough to identify a bad interval.

After the interval is divided, another BHM fit is attempted on two intervals (lines~17-26). This fit already has smaller $\chi^2_n/\tilde{n}$ values than the previous one, but still fails the threshold on several levels. Both spline intervals are then again checked separately (lines~27-31 and 32-36, respectively) and both fail the goodness-of-fit check on level~3. Note that level~0 is not present in the individual interval checks, because the bin on this level is larger than each of the spline intervals.

The intervals are numbered consecutively, but additional information is provided so that their location can be recovered (see e.g.\ lines~27 and 32). The boundaries of an interval always coincide with the boundaries of a bin on a certain hierarchy level (denoted by ``order'') and ``number'' denotes the number of this bin. 

After the intervals are again divided, the resulting BHM fit (lines~38-47) is acceptable. No separate interval checks need to be performed and the code exits with the fit result. If \verb|PrintFitInfo| is requested, the goodness-of-fit information of the final result is output again at the end. This includes the $\chi^2_n/\tilde{n}$ values on each level $n$, the unit standard deviation $\sqrt{2/\tilde{n}}$ of the corresponding $\chi^2$-distribution, as well as the number of standard deviations by which $\chi^2_n/\tilde{n}$ exceeds 1 on each level (last column). If $\chi^2_n/\tilde{n}\leq1$ the latter value is 0.

\subsection{Example 2}
\label{subsec:ex2}
This example demonstrates BHM fits for different choices of the threshold $T$. The sampled distribution is a decaying exponential (\texttt{Function=exponential}),
\begin{equation}
f(x)=\alpha\exp(-3x),
\end{equation}
normalized on the interval $[1,3]$, which implies $\alpha=3e^9/(e^6-1)$. The function is sampled on the interval $[1,2.8]$, so that there is a finite number of values $N_{\rm exc}$ sampled outside of the histogram bounds. The total number of sampled points in this example is \verb|SampleSize=100000|.

\begin{figure*}[ht]
\begin{centering}
\includegraphics[width=\columnwidth]{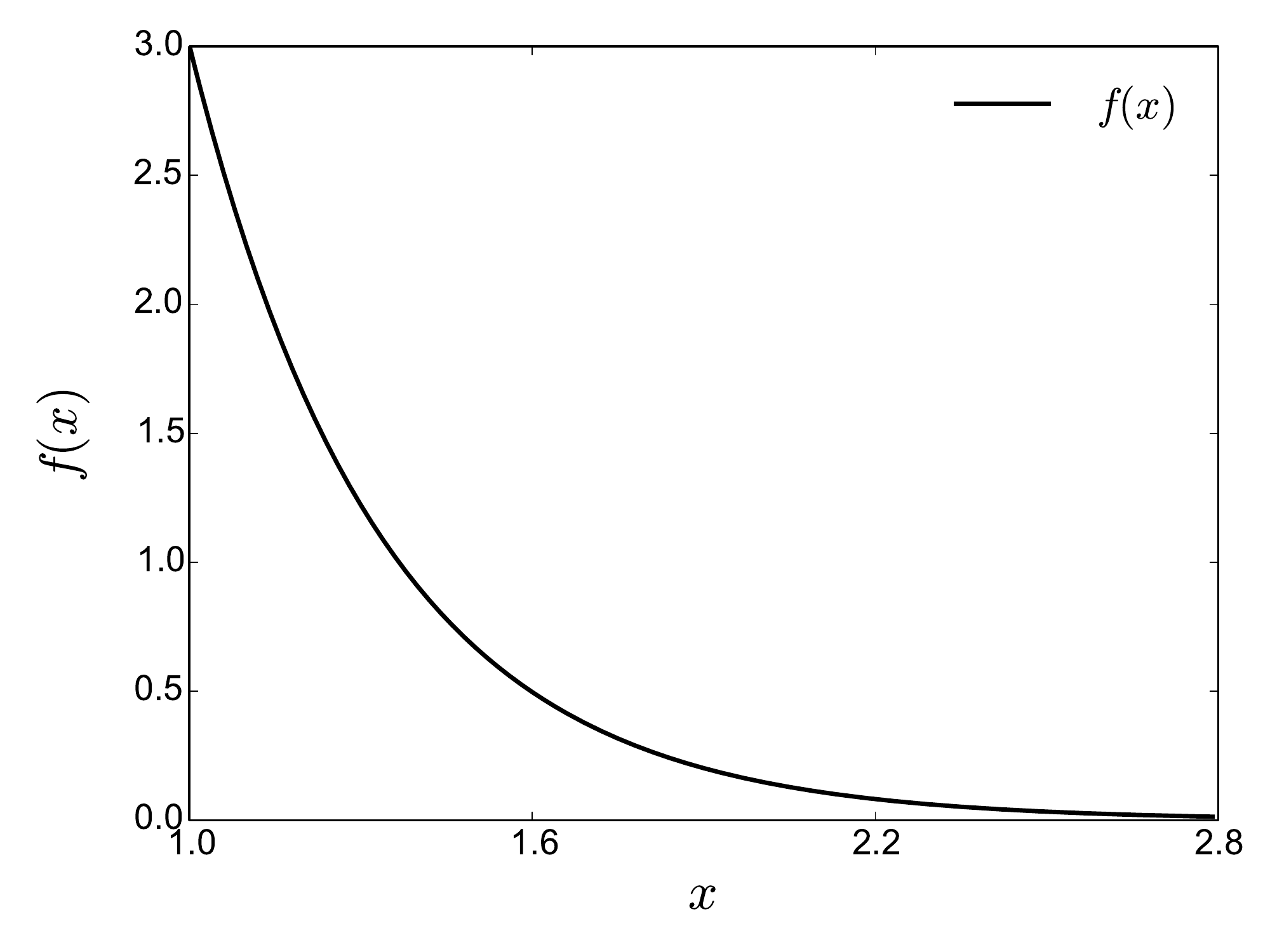} 
\includegraphics[width=\columnwidth]{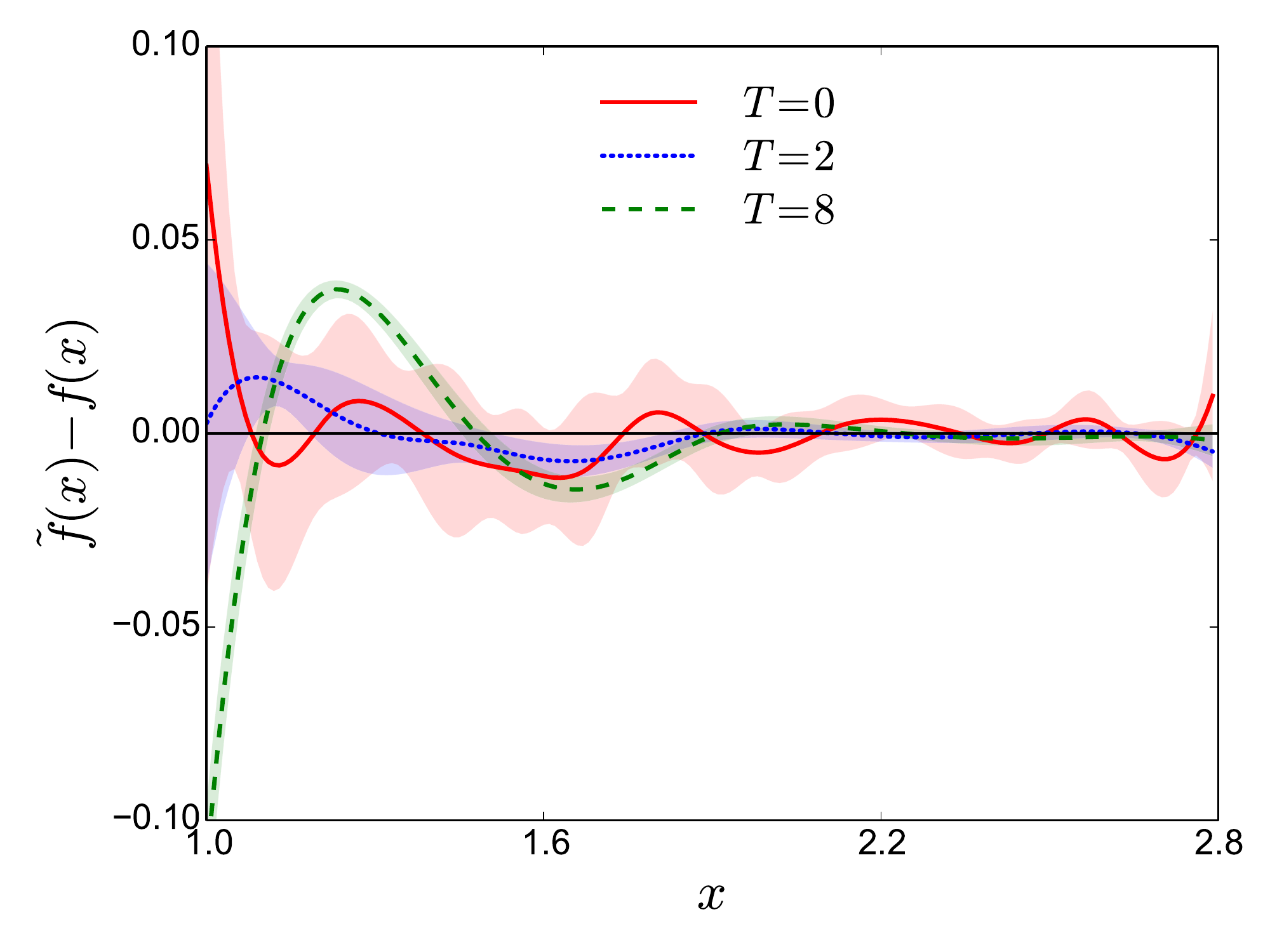} 
\caption{Decaying exponential test function (left panel). BHM fits of the test function with different goodness-of-fit thresholds (right panel).\label{fig:ex2}}
\end{centering}
\end{figure*}
The histogram data is fitted with {\BHM} using the default parameters, with the exception of the parameters defining the fit acceptance threshold, which is set to be fixed at $T=0$, 2, and 8, respectively. This can be achieved by either setting the value of \texttt{ThresholdMax} to be equal or less than the value of \texttt{Threshold}, or by setting \texttt{ThresholdSteps=0}. The fit results are shown in Fig.~\ref{fig:ex2}.

For all threshold values an acceptable fit exists, but with different interval divisions. The extremely low threshold value $T=0$ (which means that only fits with $\chi^2_n/\tilde{n}\leq1$ are accepted) yields an overfitted spline with 12 spline pieces. The value $T=2$ produces a suitable fit with 3 spline pieces that captures the shape of the test function well. The very high value $T=8$ yields an underfitted spline with only 2 pieces. This spline deviates strongly from the true function and the error on the spline is severely underestimated.

\subsection{Example 3}
This example demonstrates that \code{BHM} works for both uniform and non-uniform input histograms. The sampled distribution, 
\begin{equation}
f(x)=0.2G(0,0.2)+0.4[G(2,1)+G(-2,1)],
\end{equation}
is a linear combination of three Gaussians $G(\mu,\sigma)$ with mean $\mu$ and standard deviation $\sigma$ (\texttt{Function=triple\_gaussian}). It has several distinct features and resembles a physically relevant case. 

We sample \verb|SampleSize=1000000| data points on the interval $[-5,5]$ into a uniform and a non-uniform histogram, both with $2^8$ bins. Note that the non-uniform histogram binning is predefined and cannot be adjusted by changing the \verb|PowerBins| entry. The non-uniform histogram bins are smaller in the center of the domain (where the sampled function has a sharp feature) and increase exponentially in size towards the domain boundaries. The smallest bin size is equal to the domain length divided by $2^{12}$. The non-uniform histogram is always collected in addition to the customizable uniform histogram if \texttt{Function=triple\_gaussian} is chosen and is output into the file \texttt{nonuniform\_histogram.dat}.

\begin{figure*}[ht]
\begin{centering}
\includegraphics[width=\columnwidth]{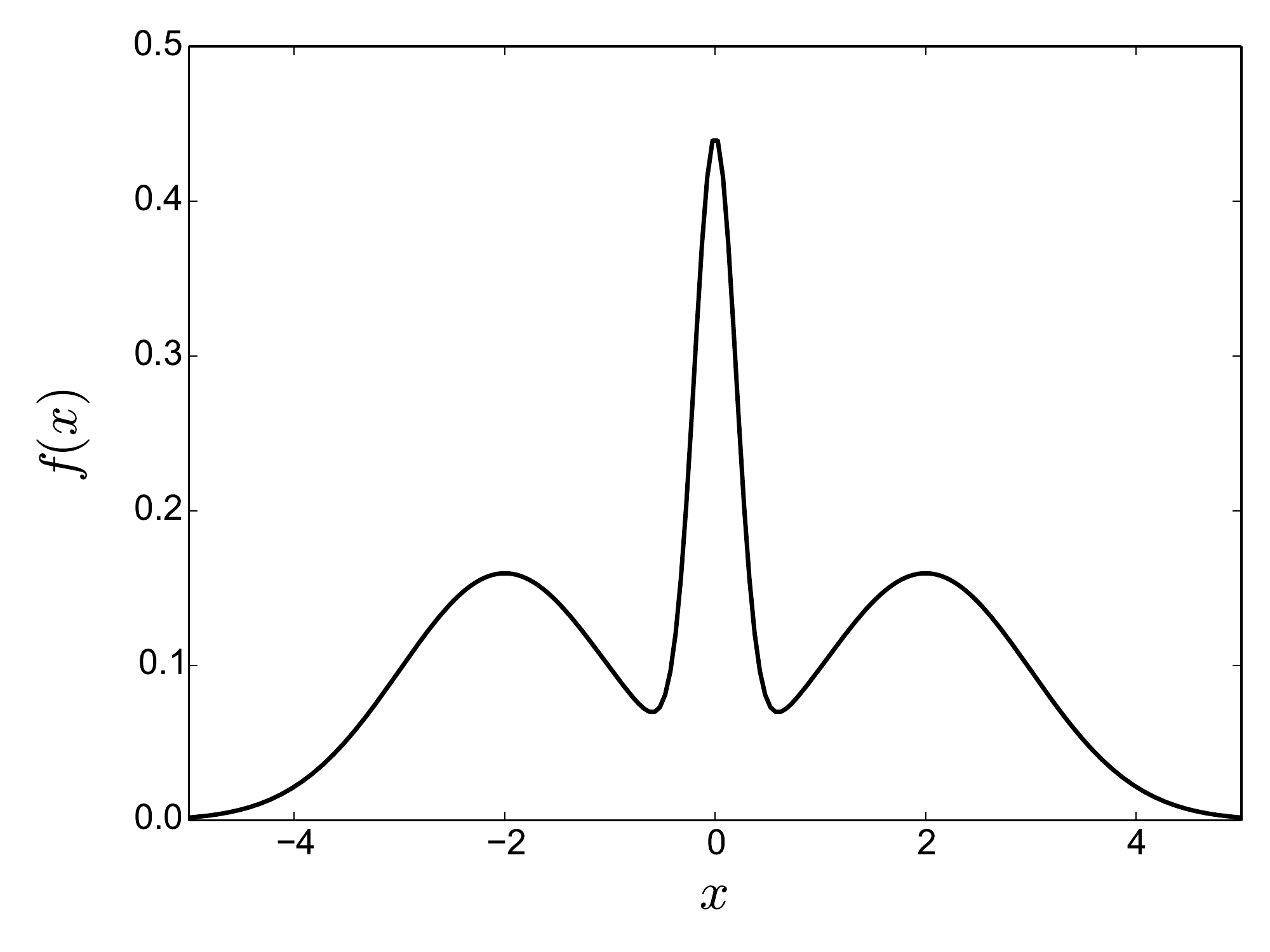} 
\includegraphics[width=\columnwidth]{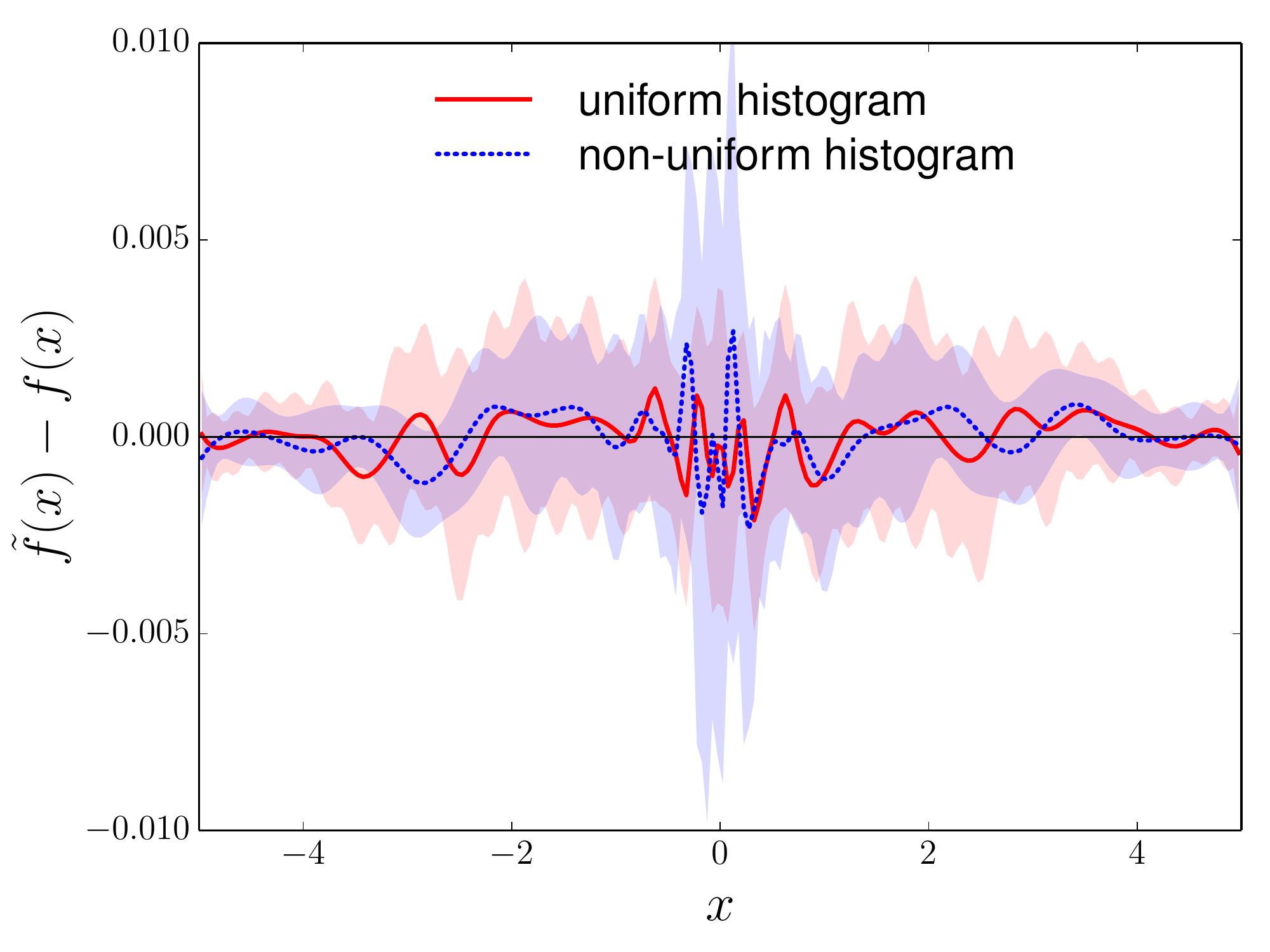} 
\caption{Triple Gaussian test function (left panel). BHM fits of the test function based on a uniform histogram and a histogram with bins of different size (right panel).\label{fig:ex3}}
\end{centering}
\end{figure*}

The fit results are shown in Fig.~\ref{fig:ex3}. Both histogram divisions produce fits of similar quality that reproduce the tested distribution well. Since \code{BHM} automatically considers combinations of elementary bins, there is no need for a case-specific implementation of a non-uniform histogram grid. Note that sampling the same data in a uniform histogram with $2^{12}$ bins produces nearly the same fit as when using $2^8$ uniform bins in this example. 

\section{Acknowledgments}
This work was supported by the Simons Collaboration on the Many Electron Problem and by the National Science Foundation under the grants PHY-1314735 (O.G., N.P., and B.S.) and DMR-1720465 (N.P. and B.S.). O.G.\ also acknowledges support by the US-Israel Binational Science Foundation (Grants 2014262 and 2016087).

\bibliographystyle{elsarticle-num}

\bibliography{bhm}

\end{document}